\documentclass{sty/llncs2e/llncs}
\usepackage{amssymb}
\usepackage{times}
\usepackage{amsfonts}

\usepackage{subfig}
\usepackage{wrapfig}
\usepackage{float}
\usepackage{multirow, graphicx}
\usepackage{algorithm}
\usepackage{algpseudocode} 
\usepackage{amsmath}

\usepackage{sty/simplemargins}
    \setleftmargin{1.750in}
    \setrightmargin{1.750in}

\usepackage{setspace} 
\usepackage{xspace}
\usepackage{hyperref} 
\usepackage{url}
\usepackage{sty/widetext} 
\usepackage{pifont}
\newcommand{\cmark}{\ding{51}}%
\newcommand{\xmark}{\ding{55}}%

\usepackage{stackrel}

\ifdefined\TTSTYLETARGETreport
\usepackage[
    backend=bibtex,
    maxbibnames=100,
    firstinits=true,
    bibstyle=alphabetic,
    citestyle=alphabetic
]{biblatex}
\fi

\usepackage{graphicx, multirow}

\usepackage{tabularx, multirow, rotating} 
\usepackage{subfig} 
\usepackage{algorithm} 
\usepackage{algpseudocode} 
\usepackage{ulem} 
\usepackage{color} 

\usepackage{listings} 
\lstdefinestyle{Oracle}{basicstyle=\ttfamily,
                        keywordstyle=\lstuppercase,
                        emphstyle=\itshape,
                        showstringspaces=true,
                        }
\makeatletter
\newcommand{\lstuppercase}{\uppercase\expandafter{\expandafter\lst@token
                           \expandafter{\the\lst@token}}}
\newcommand{\lstlowercase}{\lowercase\expandafter{\expandafter\lst@token
                           \expandafter{\the\lst@token}}}
\makeatother

\usepackage{tikz}

\newif\ifboldnumber

\algrenewcommand\alglinenumber[1]{%
  \footnotesize\ifboldnumber\bfseries\fi\global\boldnumberfalse#1:}

\begin{document}
\title{
Secure Consistency Verification for Untrusted Cloud Storage by Public Blockchains
}

\author{
Kai Li\hspace{0.25cm}
Yu-zhe (Richard) Tang\hspace{0.25cm}
Beom Heyn (Ben) Kim$\ ^\dag{}$\hspace{0.25cm}
Jianliang Xu$\ ^\ddag{}$\hspace{0.25cm}\\
\ifdefined\TTUT
\fi
}

\institute{
    \fontsize{10}{10}\selectfont\itshape
    Syracuse University, New York, USA\\
    \fontsize{10}{10}\selectfont\itshape
    $^\dag{}$University of Toronto, Ontario, Canada\\
    \fontsize{10}{10}\selectfont\itshape
    $^\ddag{}$Hong Kong Baptist University, Kowloon Tong, Hong Kong 
\ifdefined\TTUT
\fi
}

\maketitle

  \providecommand{\ssssp}{{\sc SS\_SSP}\xspace}
\newcommand{\tremark}[1]{\footnote{\textcolor{red}{(Ting's comment: #1)}}}
\newcommand{\xremark}[1]{\footnote{\textcolor{red}{(Xin's comment: #1)}}}
\newcommand{\jj}[1]{\footnote{\textcolor{blue}{(Jiyong: #1)}}}
\newcommand{\yz}[1]{\footnote{\textcolor{red}{(Yuzhe: #1)}}}
\newcommand{\yue}[1]{\footnote{\textcolor{Brown}{(Yue: #1)}}}

\definecolor{mygreen}{rgb}{0,0.6,0}
\lstset{ %
  backgroundcolor=\color{white},   
  basicstyle=\scriptsize\ttfamily,        
  breakatwhitespace=false,         
  breaklines=true,                 
  captionpos=b,                    
  commentstyle=\color{mygreen},    
  deletekeywords={...},            
  escapeinside={\%*}{*)},          
  extendedchars=true,              
  keepspaces=true,                 
  keywordstyle=\color{blue},       
  language=Java,                 
  morekeywords={*,...},            
  numbers=left,                    
  numbersep=5pt,                   
  numberstyle=\scriptsize\color{black}, 
  rulecolor=\color{black},         
  showspaces=false,                
  showstringspaces=false,          
  showtabs=false,                  
  stepnumber=1,                    
  stringstyle=\color{mymauve},     
  tabsize=2,                       
  title=\lstname,                  
  moredelim=[is][\bf]{*}{*},
}

\begin{abstract}
This work presents ContractChecker, a Blockchain-based security protocol for verifying the storage consistency between mutually distrusting cloud provider and clients. Unlike existing protocols, the ContractChecker uniquely delegates log auditing to the Blockchain, and has the advantages in reducing client cost and lowering requirements on client availability, lending itself to modern scenarios with mobile and web clients.

The ContractChecker collects the logs from both clients and cloud server, and verifies the consistency by cross-checking the logs. By this means, it does not only detect the attacks from malicious clients and server forging their logs, but also is able to mitigate those attacks and recover the system from them.
In addition, we design new attacks against ContractChecker exploiting various limits in real Blockchain systems (e.g., write unavailability, Blockchain forks, contract race conditions). We analyze and harden the security of ContractChecker protocols under these proposed new attacks.

We implement a functional prototype of the ContractChecker on Ethereum/Solidity. By experiments on private and public Ethereum testnets, we extensively evaluate the cost of the ContractChecker in comparison with that of existing client-based log auditing works. The result shows the ContractChecker can scale to hundreds of clients and save client costs by more than one order of magnitude. The evaluation result verifies our design motivation of delegating log auditing to the Blockchain in ContractChecker.
\end{abstract}

\section{Introduction}
Today, the cloud-storage consistency is a pressingly important security property. With a cloud storage service (e.g., Dropbox~\cite{me:dropbox} and Amazon S3~\cite{me:s3}), the consistency indicates how reads/writes should be ordered and whether a read should reflect the latest write (i.e., freshness). This property is exploitable to a malicious cloud provider and leads to severe security consequences. For instance, when the cloud storage hosts a public-key directory (as in certificate-transparency schemes~\cite{me:ct}), a malicious cloud violating storage consistency can return to the user a revoked public key, leading to consequences like impersonation attacks and unauthorized data access. 
Making a trustworthy assertion about cloud storage consistency is fundamental to supporting information-security infrastructure in the clouds, such as public-key directories, software-update registry, personal device synchronization, etc.

Asserting the storage consistency in a client-server system entails two conceptual steps: W1) establishing a globally consistent view of operation log across clients (i.e., operation log attestation), and W2) auditing the log for checking consistency conditions (i.e., log auditing).
To the best of our knowledge, all existing approaches including Caelus~\cite{DBLP:conf/sp/KimL15}, CloudProof~\cite{DBLP:conf/usenix/PopaLMWZ11}, and  Catena~\cite{DBLP:conf/sp/TomescuD17}, assume trusted clients and rely on them to audit the log. More specifically, they make a single party attest to the log of operations (step W1) and send the attested log to individual clients, each of which audits the log against her local operations (W2). In particular, Catena~\cite{DBLP:conf/sp/TomescuD17} is a novel scheme that makes log attestation based on the Blockchain (W1 on Blockchain), yet still uses clients for log auditing (W2 by clients).
These client-based log auditing schemes require high client availability (i.e., all clients have to participate in the protocol execution) and incur high client cost (i.e., a client needs to store the global operation log of all other clients), rendering them ill-suited for applications with a large number of low-end and stateless clients; see \S~\ref{sec:targetapp} for a list of target applications.

We propose ContractChecker, a secure Blockchain-based consistency verification protocol. Distinct from existing works, ContractChecker uniquely delegates both log attestation (W1) and log auditing (W2) to the Blockchain. Concretely, ContractChecker runs a program on the Blockchain (i.e., the so-called smart contracts) to collect log attestations from clients and the server and to audit the log there for making the consistency assertion.
Comparing existing works, our approach, by delegating log auditing to Blockchain, has the advantage in lowering client availability requirement (i.e., only active clients who interact with the cloud are required to participate in the protocol) and in minimizing client overhead (i.e., a client only maintains her own operations for a limited period of time). 

While it seems straightforward to delegate the log auditing to Blockchain as a trusted third party (TTP), putting this idea into a secure protocol and system raises the following challenges: 1) How to securely collect and audit logs from mutually distrusting clients and server (detailed in \S~\ref{sec:security:serverclient})?
2) Real-world Blockchain systems are far from an ideal TTP and can be exploited to attack a log-auditing service on Blockchain (detailed in \S~\ref{sec:security:bkc}). First, a ContractChecker party, be it a client or the server, can be malicious and forge operations in her log attestation in order to trick ContractChecker to make an incorrect assertion about the storage consistency. We propose to cross-check the attestations from both the server and clients to detect any mismatch and log-forging attacks. In addition, in the presence of such attacks, we propose mitigation techniques (in \S~\ref{sec:security:serverclient:mitigation}) for the ContractChecker to securely attribute the cause of the attacks such that it can make a correct assertion despite of the attacks.

Second, real-world Blockchain systems are known to be limited in terms of E1) write unavailability (i.e., a valid transaction could be dropped by the Blockchain), E2) exploitable smart-contract race conditions (i.e., incorrect contract logic can be triggered by running contract code concurrently), and E3) Blockchain forks (i.e., a Blockchain network or data structure can be forked to multiple instances). 
We propose attacks against the use of Blockchain in the context of ContractChecker. These new attacks systematically exploit the Blockchain limitations mentioned above, and they are: a selective-omission attack exploiting Blockchain write unavailability (E1), a forking attack exploiting smart-contract race conditions (E2), and another version of forking attack exploiting Blockchain forks (E3). 

We propose countermeasures to defend the Blockchain-oriented ContractChecker attacks. To prevent write unavailability (E1), 
we increase the transaction fee during client attestations to improve the chance of Blockchain accepting a transaction in a certain period of time. 
To prevent the race conditions on ContractChecker (E2), we define a small critical section in the contract 
to secure the contract execution yet without losing the support of concurrent server and client attestations.
We prevent forking through Blockchain forks (E3) by ensuring all clients and server be aware of the presence of all Blockchain forks. 

We build a system prototype of ContractChecker based on Ethereum~\cite{me:eth}. The on-chain part runs a smart contract (in Solidity) that collects log attestations from clients and the server, crosschecks them and audits the log with consistency conditions.
We envision a novel application of ContractChecker to enforce a consistency-centric service-level agreement (SLA) with cloud providers.

We conduct cost analysis and experimental analysis on our prototype. 
We evaluate the client cost of the protocol and service under YCSB workloads~\cite{DBLP:conf/cloud/CooperSTRS10}. Our results show that on Ethereum, ContractChecker results in significant cost saving on the client side, with reasonable Gas cost on running smart contract.
We also measure the gas cost of ContractChecker and present the evaluation in technical report~\cite{me:tr19cc} (due to space limits). It shows the practicality of ContractChecker to the extent of supporting $613710$ operations at the cost of $\$100$ at a real Ether price.

The contributions of this work are:

1. {\bf New use of Blockchain}: This work addresses verifying cloud consistency based on public Blockchain. Unlike existing research, the work delegates log auditing to Blockchain, which has the advantage of reducing client costs and availability requirements. Based on this new use of Blockchain, we propose the ContractChecker consistency-verification protocol (\S~\ref{sec:rw:consistency} and \S~\ref{sec:motives}).

2. {\bf Secure protocol design}: We present a  secure protocol design for ContractChecker  among distrusting clients and cloud provider. The protocol cross-checks logs collected from different parties and then conduct consistency checks. It not only can detect the log-forging attacks from malicious clients and server, but also mitigate those attacks (\S~\ref{sec:secureprotocols}).

3. {\bf New attacks}: This work presents new security attacks against Blockchain-based log-auditing protocols. These new attacks exploits the limits in real Blockchain systems  (e.g., write unavailability, Blockchain forks, contract race conditions). We propose to harden the security of ContractChecker under these proposed new attacks and conduct an extensive security analysis (\S~\ref{sec:analysis}).

4. {\bf Prototyping and evaluation}: We implement a prototype of the ContractChecker on Ethereum/Solidity. 
By experiments on private and public Ethereum testnets, we extensively evaluate the client-side cost of the ContractChecker. The result shows the ContractChecker can scale to hundreds of clients and, in this setting, save client costs by more than one order of magnitude when compared with existing client-based log auditing works including Catena. The evaluation on client cost justifies the key design choice in ContractChecker to delegate log auditing to Blockchain (\S~\ref{sec:eval}).

\section{Preliminary: Secure Consistency Verification Protocols}
\label{sec:preliminary}
\label{sec:rw:consistency}
Suppose there are multiple clients interacting with an untrusted data-storage service. The problem of storage consistency verification is to convince all the clients that the untrusted storage service conforms to a certain storage consistency model. In the existing literature (as will be elaborated on), verifying storage consistency entails two steps: storage log attestation and log auditing. In log attestation, a conceptual trusted party (log attester) makes attestation about the collection of all storage operations and their ordering. The purpose of log attestation is to commit to a globally consistent view about the  operation ordering such that the malicious server cannot ``fork'' different views for different clients (i.e., the forking attack~\cite{DBLP:conf/podc/MazieresS02}). After a fork-free global log is established, the second step can take place which is to audit the log and check the consistency conditions over the global log of operations.

Existing work specializes the above framework. Specifically, SUNDR~\cite{DBLP:conf/osdi/LiKMS04} does not assume a trusted third party (i.e., no trusted attester) but instead relies on the untrusted server making log attestations to individual clients. Without the trusted attester, SUNDR does not prevent forking attacks, but makes them eventually detectable (i.e., the so-called level of fork consistency~\cite{DBLP:conf/osdi/LiKMS04}). In a sense, SUNDR is a fork-detectable log attestation scheme without trusted third party.
Caelus~\cite{DBLP:conf/sp/KimL15} addresses the application scenario with battery-powered clients of a cloud storage service. Caelus rotates (or time-shares) the role of log attester among personal devices to present a virtual and highly-available trusted attester. Log auditing occurs on individual clients and Caelus supports the checking of both strong and weak consistency models. Similarly, CloudProof~\cite{DBLP:conf/usenix/PopaLMWZ11} requires the untrusted server to attest to the operation log and relies on the trusted clients to audit the operation history for a variety of consistency models.
Catena~\cite{DBLP:conf/sp/TomescuD17} is a unique Blockchain-based log attestation scheme, which prevents the server-forking attack by ``abusing'' the Blockchain security in no double-spending transactions. 

\label{sec:analsysisexisting}
All existing consistency-verification schemes including the above rely on trusted clients (or data owners) to audit the operation log and check consistency conditions. In this respect, ContractChecker is the first scheme that delegates log auditing to the Blockchain. The unique design of ContractChecker is depicted in Table~\ref{tab:distinct}. Blockchain-oriented log delegation brings benefits in lower client cost and availability requirements as will be elaborated in \S~\ref{sec:motives}.

\begin{table}[!htbp] 
\caption{Distinction of the ContractChecker design: \xmark\xspace means the log operation is not supported by Blockchain, but instead by clients.}
\label{tab:distinct}\centering{\small
\begin{tabularx}{0.89\textwidth}{ |X|c|c| }
  \hline
Solutions & Attestation & Auditing 
\\
 & by Blockchain & by Blockchain
\\ \hline
SUNDR~\cite{DBLP:conf/osdi/LiKMS04}, 
Caelus~\cite{DBLP:conf/sp/KimL15}, CloudProof~\cite{DBLP:conf/usenix/PopaLMWZ11} & \xmark & \xmark
\\ \hline
Catena~\cite{DBLP:conf/sp/TomescuD17} & \cmark & \xmark
\\ \hline
 {\bf ContractChecker}& \cmark &  \cmark
\\ \hline
\end{tabularx}
}
\end{table}

\ifdefined\TTUT
\subsection{Preliminary: Caelus and Catena (Fork)}

In Caelus, a server sends an attested log to an attester who signs one attestation in one epoch. The server then broadcasts signed attestation to all clients, who will verify/audit the attestation against their local log. In Caelus, it is crucial that an attestation is not forked, as forking different attestations for verification at different clients' sites is an effective way to conceal inconsistency. No forked attestation is assured in Caelus that there is only one known attester.

Catena delegates the Caelus attester to Blockchain. Like in Caelus, it is crucial to assure that the server does not fork attestation to the Blockchain. This can be done by having trusted clients to download the 100GB Blockchain and to check no double attestations on the same epoch. 

Catena makes this no-fork checking on Blockchain more efficient. In Catena, the clients do not download 100GB Blockchain, instead, they download the history of Catena transactions. No fork can be checked by any Catena transaction fully spending its predecessor transaction. In other words, there is no double-spend transaction in 100 GB Blockchain. 
\fi

\section{Research Formulation}

\subsection{Target Applications}
\label{sec:targetapp}

The target application scenarios of ContractChecker are characterized by the following properties:
S1) The clients are of limited capability in computing, storage, and availability. This motivates them to outsource data storage to a more powerful cloud. 
S2) Violating storage consistency leads to security consequences. 
S3) The data load is low (e.g., typically lower than tens of operations per second). In particular, the low throughput properties of these typical application scenarios make it amenable for the use of Blockchain, which is known to have limited throughput in ingesting transactions.

In real world, there are many application scenarios that fit the above paradigm. As an example, consider a DockerHub~\cite{me:dockerhub} style container registry which distributes software patches for mobile apps. In this scenario, the clients are low-power smart phones (S1). Distributing a stale software image with unfixed security bugs leads to vulnerability on users' phone (S2). In terms of workloads, an IBM report~\cite{docker_fast18}  shows that among seven geo-distributed registry deployments, the busiest one serves only 100 requests per minute for more than $80\%$ of time(S3). We believe there are many real-world applications that meet our target scenarios, ranging from Google's certificate-transparency logs~\cite{me:ct}, iCloud style personal-device synchronization~\cite{me:icloudsync}, Github-hosted secure software development~\cite{DBLP:conf/uss/Torres-AriasACC16,me:github}, etc.

\subsection{System and Security Model}

Our system model consists of three parties: a storage server in the cloud, multiple concurrent clients and the Blockchain. Clients submit storage operations (i.e., data reads and writes) to the storage server. The operations are witnessed by the Blockchain via our ContractChecker protocol. 

\subsubsection{Clients}

Clients are data owners who outsource the data storage to the cloud. They submit operations to the cloud storage to read or update their data. Different clients' operations may occur concurrently, that is, the cloud service may process multiple operations in parallel, with their time intervals overlapped. 
A client can be active when she has submitted operations recently (we will explain the definition of active clients later) and inactive when she goes offline without any operations on the cloud.

Clients in our model are \emph{stateless} in the sense that a client does not persist state across active sessions. That is, a client does not need to remember her past operations when she is engaged with a cloud. In addition, among clients, there are no direct communication channels. That is, other than the cloud or Blockchain, clients do not communicate out-of-band. We believe this model reflects low-power clients in many real-world situations, such as smart-phone clients running web browsers. Web clients are stateless and do not directly communicate among themselves.

We assume each ContractChecker client is identified, say by her public key. When the Blockchain needs to identify a client, the client identity can be obtained and verified using client certificates issued by an offline trusted certificate authority (CA). Existing key-transparency systems such as CONIKS~\cite{DBLP:conf/uss/MelaraBBFF15} can be used here in companion with ContractChecker. 
More details about client identity management are described 
in Technical Report~\cite{me:tr19cc}.

We assume clients share synchronized clocks by running NTP protocols. Synchronized clocks will be useful in our protocol when clients are required to log operations. The accuracy of NTP protocols may affect the precision of our consistency verification results. Existing protocols can achieve highly synchronized clocks and limit clock skews to the level of milliseconds~\cite{DBLP:journals/tocs/CorbettDEFFFGGHHHKKLLMMNQRRSSTWW13}, which we believe are sufficient in our system.

\subsubsection{A Cloud Storage Service}
\label{sec:servertrust}
The cloud service hosts a data store shared among multiple clients. The service accepts clients' requests and processes them concurrently. That is, different operations from clients may be processed in parallel and in an out-of-order fashion.

Under this execution model, we consider strong consistency or linearizability~\cite{DBLP:journals/toplas/HerlihyW90}. Linearizability considers independent access to different data records and is sufficiently strong to capture necessary conditions in many security scenarios~\cite{DBLP:conf/sp/KimL15,DBLP:conf/usenix/PopaLMWZ11}. Weaker consistency that is widely adopted in modern cloud services~\cite{DBLP:journals/cacm/Terry13} is commonly less relevant to security-sensitive and critical applications. In this paper, we do not consider database serializability or isolation under multiple keys~\cite{DBLP:books/aw/BernsteinHG87}. 

We assume the cloud service makes a consistency-centric service-level agreement (SLA)~\cite{DBLP:conf/sosp/TerryPKBAA13} with the clients which states where the cloud service promises to guarantee strong consistency (more specifically, linearizability as will be defined next) over the operations it will serve. An honest server will enforce the promised consistency conditions during operational hours. 

A malicious storage server will violate the consistency conditions (as will be elaborated on), by returning a stale result for a read. In addition, a malicious server will not follow our ContractChecker protocol and tries to conceal the operation inconsistency from clients.
Consistency violation, if left undetected, can lead to severe security consequences such as impersonation attacks (recall \S~\ref{sec:targetapp}).

Moreover, we assume the cloud service is rational. While a malicious cloud server may forge operations to make an inconsistent operation history look like consistent (i.e., concealing the inconsistency), the server will not attempt to make a consistent operation history look like inconsistent. We believe this assumption reflects practical situations with SLA where the cloud server will be charged in case of verified inconsistency, and thus does not have the incentive to forge an inconsistent history.

\label{sec:cons}
{\bf Consistency definition}:
We consider the storage service exposes a standard key-value API, where each data record is a key-value pair and each storage operation accesses the target record by the data key.
A storage operation, be it a read or write, is specified by two timestamps, which respectively represent 1) the begin time when the request of the operation is sent by the client, and 2) the end time when the response of the operation is received by the client. Formally, given a key-value pair $\langle{}K,V\rangle{}$, a read operation is $V=r_{[t_b,t_e]}(K)$ and $\texttt{ACK}=w_{[t_b,t_e]}(K,V)$. Here $r/w$ denotes read or write operation. $t_b<t_e$ and they are the begin and end timestamps. If two operations are concurrent, their time intervals $[t_b,t_e]$ may overlap.

An operation history is linearizable if the following two conditions are met: 1) All operations can be mapped to a total-order sequence that is compatible with the real-time intervals of the operations. 2) There is no stale read on the total-order sequence. That is, any read operation should return the record that is fresh on the total-order. This property is also called read freshness.

Any linearizable operation history can be represented by a totally-ordered operation sequence. We denote an operation sequence by operation indices.
For instance, in $w_1w_2r_3[w_2]$, the subscription is the operation index in the total-order (described below).
Linearizability specifies operations of the same data key; for simplicity, we omit data key in this notation.
The square bracket of a read indicates the result record. 
$r_3[w_2]$ denotes a read operation ordered as the third operation in the total-order sequence and which returns the record written by write $w_2$. 

{\bf Network}: In this work, we assume a reliable network among clients, the server and Blockchain nodes. When one party sends a message to the network, the network will deliver the message to the receiver with negligible delay (comparing with the period our protocol runs). We do not consider network faults or partition in this work.

\subsubsection{Blockchain}
\label{sec:prel:bkc}
The Blockchain in our protocol is a public, permissionless Blockchain running over a large P2P network and supporting smart contract execution. Real-world examples include Ethereum and the latest version of Bitcoin. In this setting, we assume the honest majority among Blockchain miners. We believe this is a reasonable assumption as in practice there are no successful $51\%$ attacks on Bitcoin or Ethereum.

The Blockchain in our protocol is parameterized by the following arguments: block time $B$, which is the average time to find a block in a Blockchain; 
and transaction-validation delay $P$, which is the time between a pending transaction enters the memory pool and when it leaves for transaction validation; and finality delay $F$, which is the number of blocks needed to be appended after a finalized transaction. That is, a transaction is considered to be finalized in Blockchain when there are at least $F$ blocks included in the Blockchain after the transaction.

The Blockchain supports the execution of smart contracts. The contract is executed across all Blockchain miners. It guarantees the execution integrity and non-stoppability. That is, the Blockchain truthfully executes a smart contract based on the provided arguments, despite of the attacks to subvert the Blockchain (as described below). Once the execution of a contract starts, it is hard to stop the execution or to abort. 

{\bf Blockchain write availability}: The Blockchain may drop transactions based on its current load and transaction fee. A transaction with a higher fee will have a lower chance of being dropped~\cite{DBLP:conf/sp/TomescuD17} and have shorter latency to be included. This assumption will be tested in our evaluation.

{\bf Blockchain attacks}: The Blockchain is subject to a series of threats. In this work, we focus on practical Blockchain threats and exclude theoretic threats (e.g., $51\%$ attacks, selfish mining, etc.) and off-chain threats (e.g., stealing wallet coins and secret keys). 1) The Blockchain may be forked permanently due to software update as in the case of Bitcoin cash (Blockchain forks). 2) Smart contracts may contain security bugs that can be exploited by a vector of attacks~\cite{me:contractattacks}, such as reentrancy attacks, buffer overflow attacks, etc. In practice, the DAO (decentralized autonomous organization) incidents are caused by this attack vector.

\subsection{Goals}

\subsubsection{Security Goals}

In our system, there are two main threats: 1) In the case that inconsistency occurs, the malicious server wants to hide the inconsistency from victim clients. 2) In the case that all operations are consistent, a malicious client may want to accuse the benign server of the storage inconsistency that does not occur. We exclude other cases as they are not rational. For instance, we do not consider that a rational server will forge operations such that a consistent operation history will appear inconsistent. We also do not consider that a victim client will want to hide an inconsistent operation history. Due to this reason, we assume clients and the server will not collude to either hide inconsistency or accuse of false consistency.

Our security goals are listed below:

{\bf Timely detection of inconsistency against malicious server}: A malicious server cannot hide the occurrence of inconsistent operations from victim clients. To be concrete, when inconsistency occurs, the protocol will assert there are inconsistent operations in a timely manner, even when a malicious server can forge a seemingly consistent log. In addition to that, the protocol will present a verifiable proof of the inconsistency, so that it can penalize the cloud service for violating the consistency.

{\bf No false accusation of inconsistency against malicious clients}: A malicious client cannot falsely accuse a benign server of inconsistency that does not occur. Given a consistent operation history, the protocol will assert it is consistent, even when there are malicious clients who want to forge inconsistent operations in the log.

\subsubsection{Cost Goals}

{\bf Client cost}: Clients in our protocol can remain stateless. That is, running the ContractChecker protocol, a client does not need to store her operations indefinitely and can truncate operations after use (i.e., stateless). In addition, the protocol data stored on a client is limited to the client's local operation; a client does not need to store other clients' operations (i.e., local operations). These two requirements make clients lightweight and render the protocol applicable to scenarios with low-power clients.

\subsubsection{Non-goals}

{\bf Data authenticity}: Data authenticity states a malicious cloud service cannot forge a client's data without being detected. We assume an external infrastructure in place that ensures data authenticity, such as MAC or digital signatures. To be more specific, the client who writes a record signs the record, and the signature can be verified to guarantee the data authenticity of the record. Here, clients' public keys are securely distributed by a PKI or a secure communication channel.

{\bf Collusion}: In ContractChecker, we aim at security guarantees against a malicious cloud server 
\emph{or} malicious clients. However, we do not consider the case that a server colludes a client, because this would make it impossible for any third-party to detect the attack. To be concrete, a colluding client can lie about her log (e.g., omitting an operation), and the server can do the same. With the client and server both lying about their operations, any third-party cannot detect the existence of the lie. If an operation can be forged or omitted in a log, the consistency assertion over the log cannot be trusted. For instance, omitting $w_2$ in $w_1w_2r_3[w_1]$ leads to incorrect consistency assertion. 
Due to this reason, we exclude from our threat model the collusion among clients and server. 
In practice, we believe server-client collusion is a rare situation.


\ifdefined\TTUT

\subsection{Research Goals}


{\bf Problem statement}: This work aims at providing consistency guarantees of a remote low-throughput storage system, 1) with the security against a malicious server, 2) at a minimal cost to the clients, and 3) by a small cost of Blockchain. To be more specific, the security of our scheme can detect any inconsistency behavior of the malicious server. 
The cost of a client should be proportional to the size of the client's local operation history, and is independent of the global operation history.
The cost of the Blockchain should be as low as possible for the practical and cost-effective use of the scheme. 

{\bf Assumptions and non-goals}: 
1) We assume our data model contains only non-identifiable information (e.g., operation type, time, data key being accessed) which can be safely published to the Blockchain. If there is identifiable information, it is hashed before being put into key-value records. By this means, we provide the basic level of user privacy. The privacy under linkage attacks and side-channel attacks~\cite{DBLP:conf/sp/NarayananS09,DBLP:journals/corr/abs-0903-3276} is out of the scope. 
2) We assume the storage throughput in the target storage is lower than that of  Blockchain. The system administrator is assumed to have sufficient funds to purchase enough storage space on Blockchain. This setting should admit various real-world applications as described in \S~\ref{sec:targetapp}. 
We do not aim at improving the Blockchain throughput in this paper. 
3) This work focuses on logical flaws of smart contracts. We assume the contracts we developed is either free of programming bugs or can be effectively debugged by external tools (e.g., following programming patterns~\cite{me:contractpattern}).
\fi

\section{The Security Protocols}
\label{sec:secureprotocols}

\subsection{Design Motivation}
\label{sec:motives}
The key design in ContractChecker is to use Blockchain to audit the log in a client-server system (recall Table~\ref{tab:distinct}). This use of Blockchain is motivated by our observations below:

{\bf 1) Low client cost}: By delegating log audit to the Blockchain, one can significantly relieve the burden from the clients. Existing consistency-verification protocols are based on client-side log auditing~\cite{DBLP:conf/sp/TomescuD17,DBLP:conf/sp/ZyskindNP15}, which requires a client to persist not only her own operations but also operations of all other clients, 
limiting the system scalability. By delegating log auditing to the Blockchain, a client can be relieved from accessing other clients' operations, and can have a minimal overhead of receiving log-auditing results from the Blockchain.

{\bf 2) Freshness with stateless clients}: Consistency verification needs to access historical operations. With stateless clients who do not communicate, it is impossible to guarantee the freshness of any global state without trusted third party~\cite{DBLP:conf/osdi/LiKMS04,DBLP:conf/podc/MazieresS02}. The presence of Blockchain as a trusted third party is necessary and beneficial in guaranteeing the freshness of serving historical operations in the presence of server rollback attack.

{\bf 3) Security without client availability}: 
Intuitively, Blockchain serves as a highly available\footnote{The condition that Blockchain is highly available is described in \S~\ref{sec:prel:bkc}.} trusted third party, and it can be used to support clients with low availability. More specifically, in existing client-based auditing schemes, all clients are assumed to be highly available in the sense that they require inactive clients to participate in log auditing in every epoch. In a world without a highly available third-party, this requirement is necessary because otherwise it is hard to distinguish an inactive client who does not report (the benign case), and an active client whose report is suppressed by the malicious server (the attack case). With a Blockchain serving as a reliable broadcast channel among clients,\footnote{There are situations that consistency verification protocols require client-client communication, such as disseminating the global list of operations for client-based auditing and reporting attack incidents.} it can afford that an inactive client does not need to participate in the protocol execution. The reliable Blockchain makes it possible to collect all active clients' logs in the presence of untrusted server. (See \S~\ref{sec:analysis:correct} for a detailed correctness analysis)


In the following, we first present the protocol overview and then justify the design by describing the alternative designs we have explored (and dismissed). After that, we will present the details of our protocol execution.

\subsection{Protocol Overview}
\label{sec:overview}

ContractChecker is a protocol that runs among clients, a storage server and the Blockchain. The purpose of the protocol is to present a trustworthy assertion about the consistency of the operations among clients and the server. At a high level, the protocol works by asking clients and the server to periodically attest to their views of operations to the Blockchain where different views are crosschecked and consistency conditions are verified. The purpose of the crosscheck on Blockchain is to detect and mitigate the attack from a malicious server (or client) who forges her view of the log.

In our protocol, the interaction between the Blockchain and the clients/server off chain can be described by 
the following three functions. Note that the protocol runs in epochs and here we consider the operations in one epoch.

\begin{itemize}
\item
$S.\texttt{attestServerLog}(Ops_S, sk_S)$: Server $S$ declares a total-order over the global history of operations in the current epoch ($Ops_S$). She attests to the declared total-order sequence by signing it with her secret key $sk_S$. 

\item
$C.\texttt{attestClientLog}(Ops_C, sk_C)$: An active client $C$ attests to her local operations in the current epoch ($Ops_C$) by signing it with her secret key $sk_C$. Operations of a single client can also be concurrent and a client does not need to declare a total-order over her own operations. If a client is inactive, meaning she did not submit any operation in the epoch, she does not need to call this function. 

\item
$C.\texttt{consistencyResult}() = \{Y, N\}$: A client $C$ can check consistency of the current operation history. With Blockchain, the checking result is only valid when the attestations are finalized on the chain. This function is blocking until the finality is confirmed. If the attestations are not finalized, it will retry by calling \texttt{attestClientLog}.
\end{itemize}

\subsection{Protocol Design Alternatives}
\label{sec:alternative}

{\bf No server attestation}: 
Our protocol (as presented above) requires to collect the log from both clients and the server.
An alternative design is to collect the operation history only from clients. In this scheme, all clients send their local operations to the Blockchain without collecting the server log. The union of client logs can reconstruct all operations and their relations in real time (i.e., serial or concurrent with each other). 
However, there are two limits of this approach. First, the union of client logs does not have the total-order information (among concurrent operations) which is necessary to determine the storage consistency (Recall the consistency definition in \S~\ref{sec:cons}). Relying on the smart contracts to ``solve'' the operation total-order, it would be expensive in terms of gas cost. Second, having attestations from both clients and server is helpful to determine which party is lying in the presence of malicious client and server. In particular, without the server attestation, it would be impossible to mitigate attacks from malicious clients. Therefore, our protocol requires the (untrusted) server to declare the operation total-order and send her attestation to the Blockchain. 

{\bf Synchronizing public contract functions}: 
Our protocol implements log attestation as public contract functions (i.e., \texttt{attestServerLog()}, \texttt{attestClientLog()}) and implements the log crosschecking and auditing as private contract functions (i.e., \texttt{crosscheckLog()}). Our initial design is to implement all three tasks in three public contract functions that can be called from off-chain clients.

However, a problem arises with the order in which these public functions can be executed. The order in which the calls to public contract functions are issued (issuance order) may be different from the order in which the public contract functions are executed
 on Blockchain miners (execution order). For instance, if a Blockchain client can serially submit functions calls for log attestation, crosscheck and checking, the three functions may be delivered to individual Blockchain miners in a different order, say checking, crosscheck, attestation, which will break the protocol correctness. To fix this, one can rely on clients to synchronize different contract calls, say by requiring a client only calls crosscheck until it confirms the completion (or transaction finality) of the call of log attestation.
However, this client-side synchronization is extremely slow and unnecessarily time-consuming.

Therefore, our protocol implements the log crosschecking and auditing by private contract functions that can only be called by another contract on chain. 
By this means, the serial order is ensured directly by individual miners without expensive client-side finality confirmation.

\subsection{The Protocol: Execution and Construction}
\label{sec:basicprotocol}
In this section, we present the protocol in the top-down fashion. We first describe the overall protocol execution flow involving both on-chain and off-chain parties. We then describe the construction of the protocol with the details of the various primitives used in the construction..

\subsubsection{Overall Protocol Execution}

We present the overall protocol execution. The ContractChecker protocol runs in epochs. Each epoch is of a pre-defined time interval $E$. During an epoch, clients send read/write requests to the storage server in the cloud. For simplicity, we consider a client runs a single thread.\footnote{In practice, the case of a multi-threaded client can be treated as multiple single-threaded virtual clients.} The operations are logged on both sides of clients and the server as in Figure~\ref{lst:execution}). A client logs her own operations while the server logs the operations submitted by all clients. Given an operation, a client logs the start time when the request for the operation is sent and the end time when the response of the operation is received. We assume the NTP protocol in place enables all clients to log operation time with low inaccuracy. Both clients and the server store the logged operations locally. 

At the end of the epoch, both clients and the server attest to their operation log. The server calls \texttt{attestServerLog(ops\_server,$sk_S$)} where she declares a total-order sequence over the operations and signs the log with server key $sk_S$ before sending it the Blockchain. A client calls \texttt{attestClientLog(ops\_client, $sk_C$)} where she signs the logged operations with secret key $sk_C$ and sends it to the Blockchain. The total-order in the server attestation is necessary for consistency checking (recall the definition in \S~\ref{sec:cons}). 

The smart contract running on the Blockchain receives the calls of \texttt{attestServerLog(ops\_server,$sk_S$)} from the server and \texttt{attestClientLog(ops\_client, $sk_C$)}  from all active clients. After that, it runs log verification, log crosschecking and log auditing 
which will be elaborated on in \S~\ref{sec:onchain}.
The result, namely the assertion of log consistency, is stored on chain for future inquiry. 
Clients can call \texttt{consistencyResult()} to check the consistency result. 

The consistency assertion is valid only when two conditions are met: 1) all active clients' transactions \texttt{attestClientLog(ops\_client, $sk_C$)} are finalized on Blockchain.
2) the consistency assertion (i.e., the result of running \texttt{auditLog()}) is finalized on Blockchain. To guarantee the two conditions, all active clients in an epoch are synchronized among themselves and with the Blockchain (see \S~\ref{sec:syncclients} for details).

ContractChecker is parameterized by the duration of an epoch $E$, which can be set according to application-level requirements in timeliness of consistency assurance (see details in our technical report~\cite{me:tr19cc}).

\begin{figure}[!ht]
\begin{center}
    \includegraphics[width=0.65\textwidth]{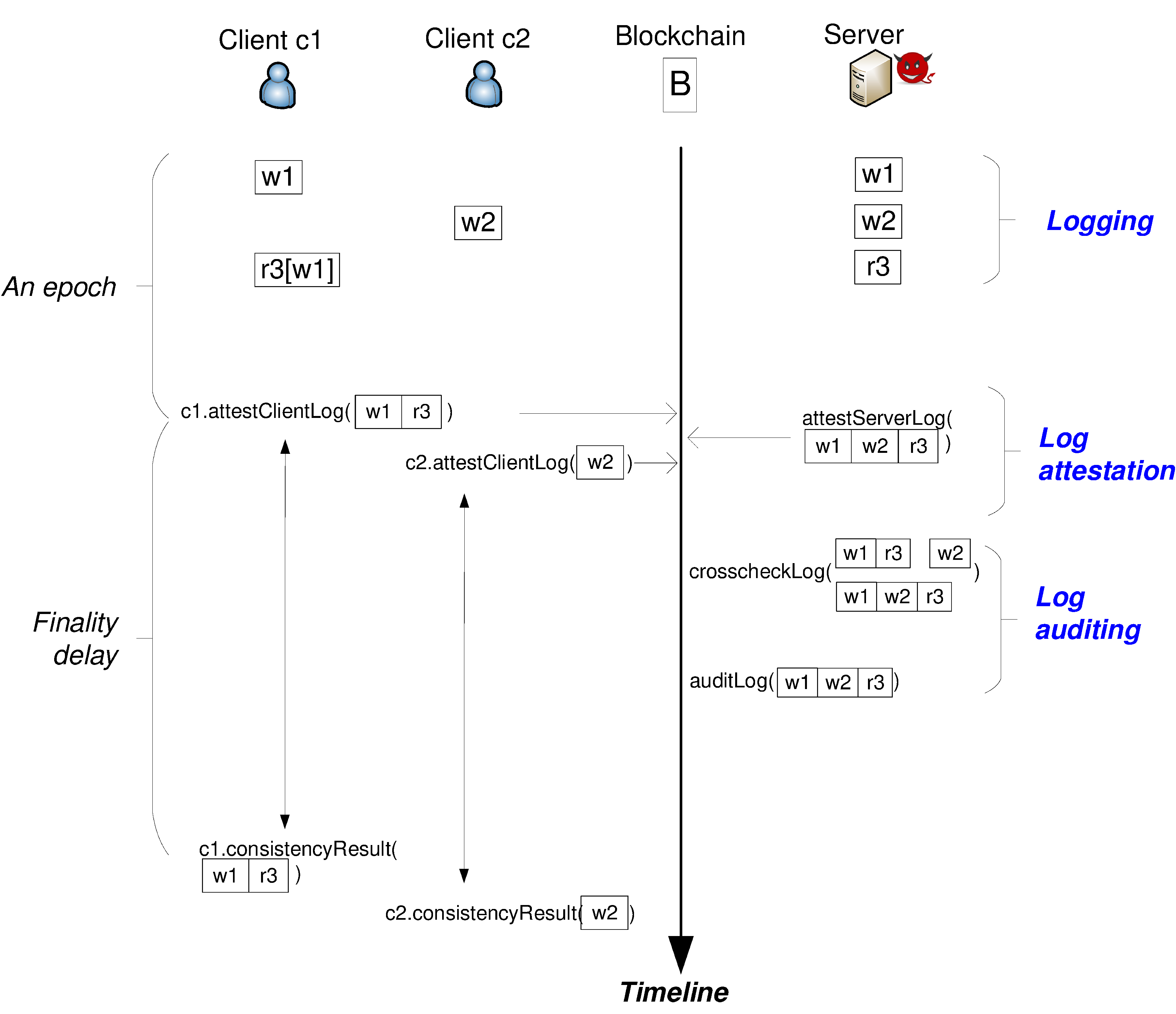}
\end{center}
\caption{Running ContractChecker: An example scenario with two clients. Client $C_1$ sends a write $w_1$ before a read $r_3[w_1]$ that returns the record written by $w_1$. Client $C_2$ sends a write $w_2$ to the server.}
\label{fig:execution}
\vspace{-0.10in}
\end{figure}

{\bf An example} is presented in Figure~\ref{fig:execution} where the ContractChecker protocol runs one epoch between two clients $C_1$ and $C_2$. An epoch can be set to multiple of Blockchain's block time. On Ethereum, it is multiply of $15$ seconds. During the epoch, client $C_1$ submits two operations to the server, namely $w_1$ and $r_3[w_1]$. $r_3[w_1]$ represents a read operation that returns write $w_1$. Client $C_2$ submits an operation $w_2$ to the server. All operations are on the same data key, and they are processed by the server in the serial order of $w_1w_2r_3[w_1]$. By the end of epoch, it first runs log attestation: 
$C_1$ calls $\texttt{attestClientLog}(w_1,r_3)$ and $C_2$ calls $\texttt{attestClientLog}(w_2)$. The server declares a total-order $w_1,w_2,r_3$ by calling $\texttt{attestServerLog}(w_1,w_2,r_3)$. 
The smart contract then stores the two client logs and one server log on the Blockchain. After that, it runs two functions: $\texttt{crosscheckLog}(\{w_1,r_3\},\{w_2\},\{w_1,w_2,r_3\})$, and $\texttt{auditLog}(\{w_1,w_2,r_3[w_1]\})$. As a result, the smart contract asserts the log is inconsistent and stores it on chain.

Meanwhile, the clients may call $C$.\texttt{consistencyResult()}. The function call will block until the transaction of $C$.\texttt{attestClientLog()} is finalized on Blockchain. On Ethereum, it takes $25$ epochs to finalize a transaction. After 25 epochs, if the transaction is finalized, the consistency assertion stored on chain can be treated as an immutable statement and be further used by applications. In addition, the client can truncate her local operations in that epoch. If the transaction is not included in the Blockchain, the client will retry $C$.\texttt{attestClientLog()}.

Now we modify the setting of this example that $w_1$ and $w_2$ are processed concurrently by the server (i.e., with overlapped time intervals). In this case, a rational server (recall \S~\ref{sec:servertrust}) will search and attest to the total-order that is consistent, namely $w_2w_1r_3[w_1]$. Note that the other total-order $w_1w_2r_3[w_1]$ also matches the real-time relation, as $w_1$ and $w_2$ are concurrent. But it contains inconsistent read $r_3[w_1]$.

Note that in this example, the ContractChecker clients are stateless (in that client $C_1$ can discard operations $w_1r_3$ and truncate the log after client attestation) and store only local operations (in that client $C_1$ does not need to store the operations of client $C_2$). This saves the client cost, comparing all existing approaches including Catena and Caelus which require clients to maintain global state without log truncation.

\subsubsection{Construction of On-chain Contracts}
\label{sec:onchain}

\definecolor{mygreen}{rgb}{0,0.6,0}
\begin{figure}[h!]
\lstset{ %
  backgroundcolor=\color{white},   
  basicstyle=\scriptsize\ttfamily,        
  breakatwhitespace=false,         
  breaklines=true,                 
  captionpos=b,                    
  commentstyle=\color{mygreen},    
  deletekeywords={...},            
  escapeinside={\%*}{*)},          
  extendedchars=true,              
  keepspaces=true,                 
  keywordstyle=\color{blue},       
  language=Java,                 
  morekeywords={*,...},            
  numbers=left,                    
  numbersep=5pt,                   
  numberstyle=\scriptsize\color{black}, 
  rulecolor=\color{black},         
  showspaces=false,                
  showstringspaces=false,          
  showtabs=false,                  
  stepnumber=1,                    
  stringstyle=\color{mymauve},     
  tabsize=2,                       
  title=\lstname,                  
  moredelim=[is][\bf]{*}{*},
}
\begin{lstlisting}
contract ContractChecker{          
  address payable ownerContract; 
  attestClientLog(Op[] ops_client, signature_client){
    if(true == verifyClientLog(ops_client, 
               signature_client, pubkey_client)){
      lock.acquire();//to prevent race conditions
      if(++attested_clients == N 
         && attested_server = 1){
         if(!crosscheckLog(ops_clients, ops_server))
           throw;
         if(!auditLog(ops_server)) throw;
      }
      lock.release();
    }
  }

  attestServerLog(Op[] ops_server, signature_server){
    if(true == verifyServerLog(ops_server, signature_server, pubkey_server)){
      lock.acquire();//to prevent race conditions
      if(attested_clients == N
         && ++ attested_server = 1){
         if(!crosscheckLog(ops_clients, ops_server)) 
           throw;
         if(!auditLog(ops_server)) throw;
      }
      lock.release();
    }
  }

  modifier crosscheckLog(Op[] ops_clients, Op[] ops_server)) returns (false) {...}
  modifier auditLog(Op[] ops_server)) returns (false) {...}
  mapping ops_server;
  mapping ops_clients;
} 
\end{lstlisting}
\caption{Smart-contract program of ContractChecker}
\label{lst:execution}
\end{figure}

The two attestations described in \S~\ref{sec:overview} invoke a smart-contract on chain. The smart contract taking as input of log attestations from the server and clients will conduct four operations: verifying client attestations (\texttt{verifyClientLog()}), verifying the server attestation (\texttt{verifyServerLog()}), crosschecking the logs of clients and server (\texttt{crosscheckLog()}), and auditing the checked log to assert consistency (\texttt{auditLog()}).

In \texttt{verifyClientLog()}, the contract would verify the client attestation using the client's public key. In \texttt{verifyServerLog()}, the contract verifies the server attestation using the server's public key. It is optional that an individual operation can be signed by both the client and server. In this case, both verification functions will verify individual operations as well using both clients' and server's public keys. Once logs are verified, the client logs are union-ed and are crosschecked with the server log to find any inequality (\texttt{crosscheckLog()}). Once the logs are successfully cross-checked, it runs log auditing (i.e., \texttt{auditLog()}) where strong consistency conditions (e.g., operation ordering and read freshness) are checked over the crosschecked server log.

\subsubsection{Finality of Client Attestations}
\label{sec:syncclients}
The finality checking is realized by all active clients staying online and confirming the finality of their own attestations. For a client to check the finality of her attestation, she simply checks if there are $F$ blocks on the Blockchain that are ordered after the block where her attestation transaction is stored. Blockchain's immutability guarantees the hardness of altering or omitting the client attestation if its finality is confirmed on the chain. If the transaction is not finalized, the client is responsible for resubmitting \texttt{attestClientLog()} 
until the confirmed finality of the transaction.

A consistency assertion is only valid when all active clients have confirmed the finality of their log attestations. Had one client's log not been confirmed on the Blockchain, a malicious server can launch the selective-omission attack (detailed in \S~\ref{sec:security:bkc:available}) that leads to an incorrect assertion. Clients wait until all clients' attestations are finalized on Blockchain (see \S~\ref{sec:security:bkc:available} for a heuristic client-synchronization scheme).

\section{Protocol Analysis}
\label{sec:analysis}

In this subsection, we analyze the protocol correctness under both benign and malicious settings. We start with the correctness with benign clients and the server.  We consider the case of lowly-available clients. We then focus on the malicious cases, by analyzing protocol security with a malicious server, malicious clients, and unreliable Blockchain.

\subsection{Correctness}
\label{sec:analysis:correct}
In ContractChecker, the protocol correctness states that if the protocol is truthfully executed by benign clients and server, a consistent operation history will be asserted as a consistent history, and an inconsistent operation history will be asserted as an inconsistent history. Informally, given any consistent operation history $Ops_S$ and any $\{Ops_C\}$ with $\cup{Ops_C}=Ops_S$, it holds that $S.\texttt{attestServerLog}(Ops_S)$, $C.\texttt{attestClientLog}(Ops_C)$, $C.\texttt{consistencyResult}()=Y$. For any inconsistency operation history $Ops_S$ and any $\{Ops_C\}$, $S.\texttt{attestServerLog}(Ops_S)$, $C.\texttt{attestClientLog}(Ops_C)$, $C.\texttt{consistencyResult}()=N$.

Analyzing the protocol correctness is straightforward. If the clients and server are benign, the protocol guarantees the authentic copies of client logs and server log are send to the Blockchain as the input of \texttt{crosscheckLog()} and \texttt{auditLog()}. The computation logic guarantees that a server log with consistency total-order will be asserted as consistency. For an inconsistency log, the server cannot find a total-order without inconsistency and the case of inconsistency will be detected by \texttt{auditLog()}.

\label{sec:clientavail}
{\bf Correctness with low client availability}: In ContractChecker, we consider active clients and inactive clients: Given an epoch, an active client is one that has submitted at least one operation to the cloud. Otherwise, it is an inactive client. ContractChecker requires only active clients in an epoch to be available to participate in the protocol by the end of the epoch. Be more precise, given an inactive client, it does not require its availability for protocol participation. 

The protocol correctness without the availability of inactive clients is straightforward. Whether inactive clients send their empty log to the Blockchain, it does not affect either the union of all client operations or the log crosscheck (\texttt{crosscheckLog()}). Therefore, an attack is detected by the server-client log inequality, which is irrelevant to the inactive clients' empty logs. 

This is in contrast with existing client-based protocols which require the availability of both active and inactive clients.
Briefly, the reason that ContractChecker does not require availability of inactive clients while other protocols do is that it cannot distinguish the benign case of an inactive client who is legitimately unavailable from the malicious case of an active client who detects an attack but whose attack report is suppressed by the untrusted server who relays the report among clients. In ContractChecker, it does not rely on the untrusted server to report the case of an attack, but instead by the trusted Blockchain.

\subsection{Security under Server/Client Attacks}
\label{sec:security:serverclient}

In this subsection, we consider the attacks launched by individual malicious clients or server. As reasoned before, our threat model excludes the collusion between a client and server. We leave the Blockchain exploits to the next subsection.

In our threat model, either a client or the server can forge her attestation to the Blockchain. Specifically, she can forge a non-existing operation (A1), omit a valid operation (A2), replays a valid operation multiple times (A3), reorders the serially-executed operations (A4). 
Recall that our system model considers a rational server who, given concurrent and consistent operations, will not declare a total-order to make it look like an inconsistent log.

A malicious server can exploit the operation forging (A1-A4) to conceal an inconsistent log and to avoid paying the penalty to victim clients. For instance, successfully omitting $w_2$ in $w_1w_2r_3[w_1]$ may fool the ContractChecker to falsely assert the operation history to be consistent. A malicious client may exploit the operation forging (A1-A4) to forge an inconsistent log and to falsely accuse a benign cloud. For instance, a client can forge an operation (A1) $w_{2.5}$ in a consistent log $w_1w_2r_3[w_2]$ to make it look like inconsistent.

ContractChecker can detect any operation forging (A1, A2) by a mismatch between server attestation and client attestation (in \texttt{crosscheckLog()}). 
Without server-client collusion, if one party, say the server, forge an operation in the server attestation, the forged/omitted operation will be found in the server/client attestation, but not in the client/server attestation.
The replay attack (A3) can be detected by multiple identical operations in the server/client log. Reordered operations (A4) can be detected by the condition that the operation order does not match the real-time order (i.e., an operation that occurs later is ordered before an earlier operation).

\label{sec:security:serverclient:mitigation}
{\bf Security hardening for attack mitigation}: ContractChecker is security-hardened to not only detect attacks but also mitigate these attacks. Mitigating the attacks requires attributing an attack to the correct cause and to recover the system from the attack for further processing. The fundamental challenge in the attack attribution is about distinguishing the cause of inconsistency between a malicious client logging a forged operation or a malicious server omitting a valid operation.

To support attack attribution and to overcome this fundamental challenge, we require each online storage operation, when it is being served, needs to be witnessed and doubly signed by both the server and the submitting client. The double signatures can establish a ground truth to distinguish the cause of attacks at the consistency-checking time. For instance, if a doubly-signed operation is found in the server attestation but not in any client attestation, then it is attributed to a malicious client omitting her operation. If a doubly-signed operation is found in a client attestation but not in a server attestation, it is attributed to a malicious server omitting her operation. Note that a malicious client forging an operation can not get away as the forged operation can not be signed by the server. Also note that we don't consider the case of server-client collision, which presents an impossibility result for attack detection. 

By this means, the ContractChecker can distinguish the cause of different attacks (i.e., A1, A2, A3 or A4) and can recover the system from attestation forging. 
The details of the attack mitigation are presented in Technical Report~\cite{me:tr19cc}.

\subsection{Security under Blockchain Exploits}
\label{sec:security:bkc}
\subsubsection{Exploiting Blockchain Write Unavailability}
\label{sec:security:bkc:available}
Recall that the practical Blockchain systems exhibit low write availability, and may drop valid transactions. Given a faulty Blockchain like this, a malicious server can selectively omit operations in her attestation such that dropped transactions of client attestations correspond to the omitted operations in the server attestation. By this means, the server can omit operations without being detected by ContractChecker, in a way to conceal inconsistency. For instance, in Figure~\ref{fig:execution}, Client $C_2$'s call of \texttt{attestClientLog($w_2$)} can be dropped by the Blockchain. 
The malicious server, observing $w_2$ is not included in the Blockchain, can selectively omit the operation in her attestation. This will allow the forged log (with omitted operations) to pass the log crosschecking (more specifically, \texttt{crosscheckLog($\{w_1,r_3[w_1]\}$,$\{w_1,r_3[w_1]\}$)}, which further tricks the ContractChecker to assert incorrectly that the operation history (which is actually $w_1,w_2,r_3[w_1]$) is consistent. Because in this attack, the server selectively omit operations based on dropped transactions in Blockchain, we call this attack by \emph{selective-omission} attack.

{\bf Security hardening against Blockchain unavailability}: The selective-omission attack can be prevented if the finality of any client's log attestation in Blockchain can be assured of. To prevent the miss of a transaction on Blockchain, a common paradigm is to resubmit the transactions. A naive resubmission policy is to resubmit every $F$ block until the transaction is finalized. With this naive policy, it may lead to an unwanted situation where the resubmitted transaction keeps being declined (e.g., due to low transaction fee). 

Enforcing a time bound on transaction finality is crucial to the security of ContractChecker. That is, ContractChecker's security relies on whether a transaction (after resubmission) can be guarantee to succeed before a pre-scribed timeout. Because if both clients and Blockchain are allowed to be unavailable in an arbitrarily long time, it is impossible to distinguish between the benign case of an inactive client where the client does not send a transaction and the malicious case of an unavailable Blockchain under selective-omission attacks.

Therefore, our security hardening technique is to enforce a time bound on the transaction finality. Instead of resubmitting transactions, we require high transaction fee (e.g., based on a heuristic, $20\cdot{}10^{9}$ Wei in Ethereum transactions) and increase the chance of transaction being accepted in the first place namely in $F$ blocks. The higher transaction fee is, the more likely a transaction is being accepted to be included in the Blockchain. The relationship between Blockchain write availability and transaction fee is evaluated in 
our technical report~\cite{me:tr19cc}
With the high transaction fee, all clients wait for $F$ blocks after their transactions. 

{\bf Analysis of attack prevention}: If all active ContractChecker clients execute the transaction resubmission policy, the selective-omission attack can be prevented. What follows is the security analysis: Assume the resubmission policy can guarantee the high chance (the evaluation is in our technical report~\cite{me:tr19cc}) 
 that all clients' transactions (for \texttt{attestClientLog()}) are included in the Blockchain before the timeout of $F$ blocks. The selective-omission attack cannot succeed because any valid operation can be found in the client log attestations and the omission of the operation in the attack can be detected by the mismatch in log crosschecking. In the previous example, by the time $w_2$ is included in the Blockchain (before $F$ blocks), the ContractChecker can detect the $w_2$ is absent in the server log as it appears in Client $C_2$'s log. The ContractChecker can identify the attack, recover the server log for making a correct assertion about the operation inconsistency.

\subsubsection{Forking Attacks}

{\bf Preliminary on forking attacks}: Consider multiple clients share a state hosted on a server. In this setup, a forking attacker is a malicious server who forks the shared state and serves different clients with different (forked) states. In client-based consistency protocols (e.g., Catena and Caelus), the malicious server may fork her log attestations for different auditing clients, such that the forked global logs appear to be consistent from different clients' local views. 

In ContractChecker, both clients and server send their attestations to the Blockchain where the crosschecking occurs. The regular forking attack where the server forks her view among clients may not succeed because the Blockchain essentially provides a ``gossiping'' channel among clients and miners. However, Blockchain provides only an imperfect gossiping channel and there are exploits one can leverage to break the gossiping channel.

The central idea here is that the server can fork the inputs to the \texttt{crosscheckLog()} function in smart contract so that the function instances running on different miners take forked inputs (from the server and clients) and falsely accept separately. The forking can be facilitated by exploits such as smart contract races and Blockchain forks. 

\label{sec:security:bkc:fork}
{\bf Exploiting Blockchain forks}, the server log is forked by the malicious server and the logs of different clients may be forked due to network partitions. This could effectively lead to a forking attack. Concretely, when the network partition separates the Blockchain network and clients into different partitions and the server is in a position to mount the forking attack and to send different partitions different server views. For instance, consider the server wants to hide a ``total-order anomaly'' sequence $w_1|w_2,r_3[w_1]|r_4[w_2]$ where $w_1$ and $w_2$ are concurrent, $r_3$ and $r_4$ are concurrent, and the four operations are submitted from four different clients  (respectively $C_1,C_2,C_3,C_4$). The network partitions clients and Blockchain at the attestation time to two partitions, say $P_1$ and $P_2$, where in $P_1$ clients $C_1$ and $C_3$ attest to $w_1$ and $r_3$ to the Blockchain, and in $P_2$ clients $C_2$ and $C_4$ attest to $w_2$ and $r_4$ to the Blockchain. Then, the server is in a position to launch a forking attack by sending a view of server log ($w_2w_1r_3[w_1]$) to Blockchain partition $P_1$ and another (forked) view of server log ($w_1w_2r_4[w_2]$) to Blockchain partition $P_2$. Because the client attestations are also forked due to network partition, the smart contracts of ContractChecker running in both $P_1$ and $P_2$ will (falsely) accept.

To make this attack successful, the key is the Blockchain fork due to the network partition. In practice, however, this type of Blockchain fork is temporary and is resolved when the network reconnects. When the network reconnects, one of the Blockchain forks will be orphaned, which will leave some client attestation omitted eventually. For instance, if the fork in $P_1$ is chosen and that in $P_2$ is orphaned, the client attestations from $C_2$ and $C_4$ can be permanently lost. In our scheme, this attack is detected by our client-side technique (\S~\ref{sec:security:bkc:available}) which will check the finality of transactions after sufficient time.

For other types of Blockchain forks, such as caused by software updates, forking attacks cannot succeed. Because client attestations are not separated, all Blockchain forks will receive the same set of attestations of all clients. To hide an inconsistent anomaly (e.g., stale read), it would require the fresh write operation to be omitted in the Blockchain, which is impossible in the presence of all client attestations. Note that by forking the total-order declaration, it is not sufficient to hide inconsistency anomaly.


\label{sec:security:bkc:race}
{\bf Exploiting smart-contract races}, the forking server attempts to update the server log in the middle of function execution of \texttt{crosscheckLog()} by exploiting race conditions. The goal is to make client logs cross-checked with different (forked) versions of server log.
Concretely, consider the function-call sequence \texttt{$S$.serverAttestLog($w_1,w_2,r_3[w_1]$)}, \texttt{$C_1$.clientAttestLog($w_2$)}, \texttt{$C_2$.clientAttestLog($w_1,r_3[w_1]$)}, \texttt{$S$.serverAttestLog($w_2,w_1,r_3[w_1]$)}. Suppose before \texttt{$S$.serverAttestLog($w_2,w_1,r_3[w_1]$)} runs, the contract already enters \texttt{crosscheckLog()} where the server log $w_1,w_2,r_3[w_1]$ is being crosschecked with  client logs $w_2$ and $w_1,r_3[w_1]$. When \texttt{$S$.serverAttestLog($w_2,w_1,r_3[w_1]$)} runs, it might happen that the call of \texttt{$S$.serverAttestLog($w_2,w_1,r_3[w_1]$)} replaces the variable of \texttt{ops\_server} \emph{during} \texttt{crosscheckLog()}. This may lead to the unwanted situation that client log $w_2$ is crosschecked with one version of server log $w_1,w_2,r_3[w_1]$, and client log $w_1,r_3[w_1]$ is crosschecked with another version of server log $w_2,w_1,r_3[w_1]$. In this situation, server log $w_2,w_1,r_3[w_1]$ will survive as the successfully crosschecked version, which will lead to an incorrect consistency.
We call this attack by \emph{forking-by-races} attack.

The ContractChecker prevents the forking-by-races attack by synchronizing the critical functions and avoiding concurrency. Concretely, in ContractChecker, we define a critical section around the functions \texttt{crosscheckLog()} and \texttt{auditLog()}, such that the execution of these two victim functions needs to be serialized with other functions. To be more specific, in the forking-by-races attack, 
the attacker (i.e., calling function \texttt{$S$.serverAttestLog($w_2,w_1,r_3[w_1]$)}) and the victim (i.e., running function \texttt{crosscheckLog()}) are forced to execute in a serial order. Without concurrent execution of attacking and victim functions, the attack cannot succeed.

\subsection{Incentivize DoS Attacks}

{\bf DoS attacks}: With the policy of high transaction fee, it is tempting to design the following DoS attack: An attacker can 1) join the Blockchain network as an honest miner who mines and executes the smart contract to collect transaction fee, and 2) join the ContractChecker protocol as a client to send transactions and run DoS attacks. The rationale behind the attack is that the attacker can make coins from her ``honest miner'' role and spend the coins to generate some high-fee transactions sent to the Blockchain. This effect of the high-fee transactions is to race against the transactions sent from honest ContractChecker clients and to prevent them from being included in the Blockchain. The honest clients, in the hope of including their transactions in the Blockchain, would be forced to increase their fee again, which will be collected by the DoS attacker miner and inventive them to launch more DoS attacks.

We believe the above attack is hard to succeed in existing public Blockchain which runs in a large P2P network. The chance of making a perceivable amount of coins (by mining and running smart contract) in today's public Blockchain is slim. A self-sustainable DoS attack would need a large investment (similar to $51\%$ attack) to bootstrap. Without it, any DoS attacker would lose coins and cannot sustain.

\section{Evaluation}
\label{sec:eval}
\label{sec:macro}

In this section, we evaluate the client cost in ContractChecker.
We first present an analysis of client cost and then present our experimental results. 

\subsection{Cost Analysis}
{\bf Cost model}: For a consistency verification protocol, either client-based schemes or ContractChecker, its execution can be modeled by the following: Clients periodically send their log attestations and check the consistency results. In this process, a client's cost is characterized by 1) how many operations the client needs to store, and 2) how long the client needs to maintain an operation. In our cost model, we accredit to one cost unit storing one operation by a client in one epoch. Given a process of $T$ epochs, a client's total cost is the sum of cost units of all operations stored in the client in all epochs. If an operation is stored continuously in a client for $T$ epochs, it is counted as $T$ units. 

Based on the above model, we present a cost analysis of client-based auditing schemes and ContractChecker. 

{\bf Client cost in client-based auditing}:
In client-based auditing schemes, a client needs to access operations submitted by all clients in the current epoch, including herself and other clients; these operations may only need to be stored by the clients in one epoch, assuming a trusted third-party attester. 
For simplicity, we omit the cost of accessing historical operations. Instead, we focus on a low-bound estimate of the client cost in client-based auditing schemes:

\begin{equation}
CCost_{ClientAudit} \geq{} T\cdot{}M\cdot{}N
\label{eqn:clientaudit0}
\end{equation}

The above equation considers that the protocol is run in $T$ epochs among $N$ clients where each client in one epoch submits $M$ operations on average. The equation reports the total cost. We focus on the ``unit'' client cost, that is, the client cost per epoch and per operation:

\begin{equation}
CCost^{Avg}_{ClientAudit} =
CCost_{ClientAudit}/(T\cdot{}M) 
\geq{} N
\label{eqn:clientaudit}
\end{equation}

The above equation considers a client's cost in a process of running client-based auditing scheme in $T$ epochs, with totally $N$ clients where on average each client in one epoch submit $M$ operations. 

{\bf Client cost in ContractChecker}: 
In the above setting, a ContractChecker client only needs to store her own operations, that is, $M$ operation per epoch. 
But the ContractChecker client needs to keep an operation for, instead of the current epoch, but an extended period of time denoted by $F_t/E$ epochs. $F_t$ is the average delay for successfully submitting a client attestation to the Blockchain in ContractChecker. Specifically,

\begin{equation}
F_t=r\cdot{}(B\cdot{}F+P)
\label{eqn:ft}
\end{equation}

Here $r$ is the average times to resubmit a client attestation, and $F$/$B$/$P$ is the finality delay/block time/average wait time for transaction validation. Thus, ContractChecker's client cost is as follows. Note that in ContractChecker, log auditing is fully delegated and clients are relieved from accessing historical operations.

\begin{equation}
CCost^{Avg}_{ContractChecker} = F_t/E
\label{eqn:cc}
\end{equation}

Therefore, the cost saving of ContractChecker comparing existing client-based auditing schemes is:
$\frac{CCost^{Avg}_{ClientAudit}}{CCost^{Avg}_{ContractChecker}}
\geq \frac{E\cdot{}N}{F_t}$.
This formula shows that the cost saving of ContractChecker comparing existing work depends on three factors: the total attestation delay $F_t$, the maximal number of clients $N$, and epoch time $E$.
The larger $E$ is, the more costs the ContractChecker saves. Because with a larger $E$, the client-based schemes have to store an operation for longer time. The larger $F_t$ is, the fewer costs the ContractChecker saves. Because with a larger $F_t$, an operation needs to be stored on a ContractChecker client for longer time. The more clients there are (a larger $N$), the more costs the ContractChecker saves. Because with a larger $N$, the more operations a client needs to store (in client-based auditing schemes).

\subsection{Experiments}
\subsubsection{Client Cost}

\begin{figure*}[!hbt]
\begin{center}
\subfloat[With varying number of clients]{
   \includegraphics[width=0.225\textwidth]{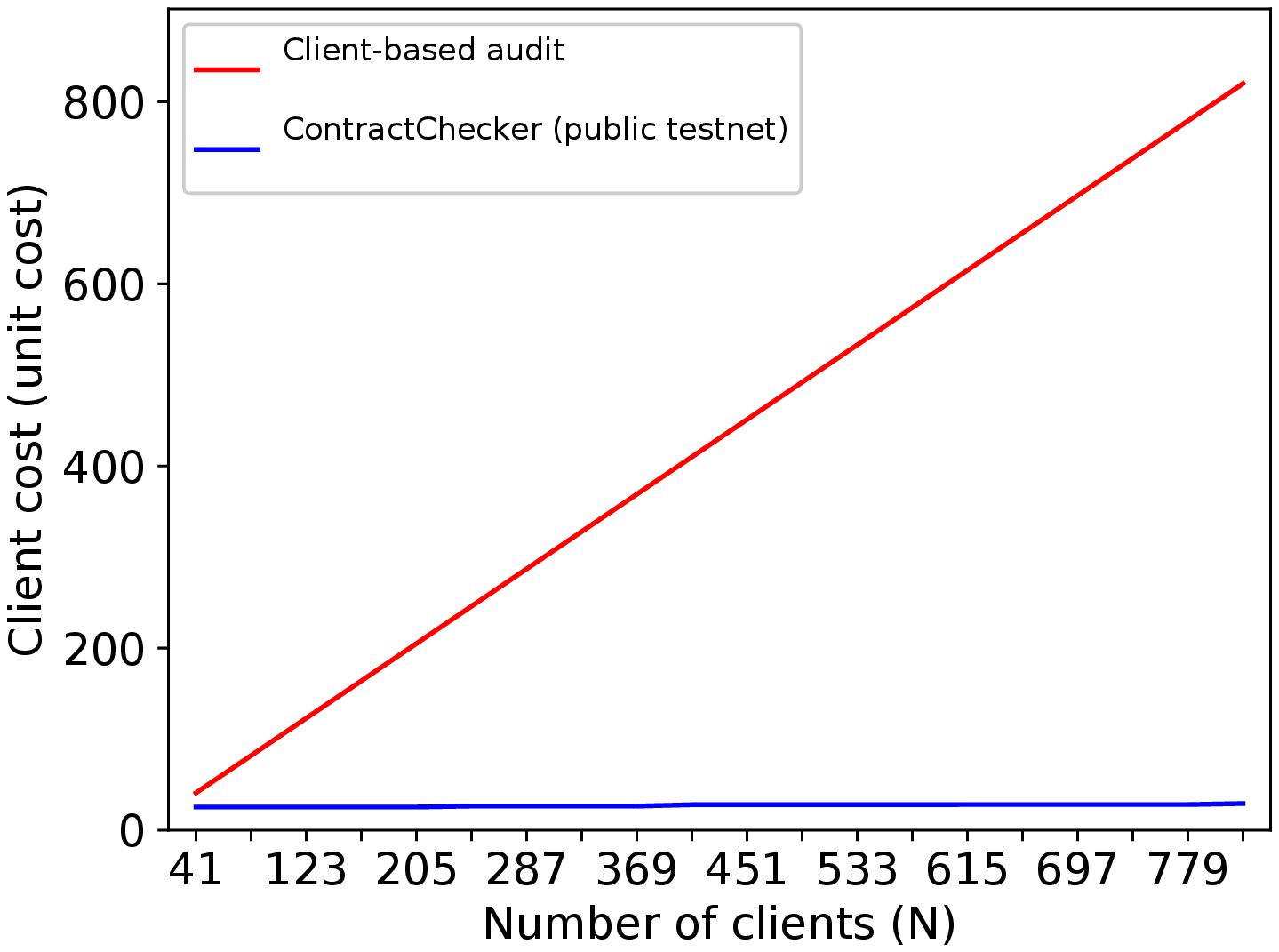}
    \label{exp:numclient:1}
}
\subfloat[With varying number of clients]{
   \includegraphics[width=0.225\textwidth]{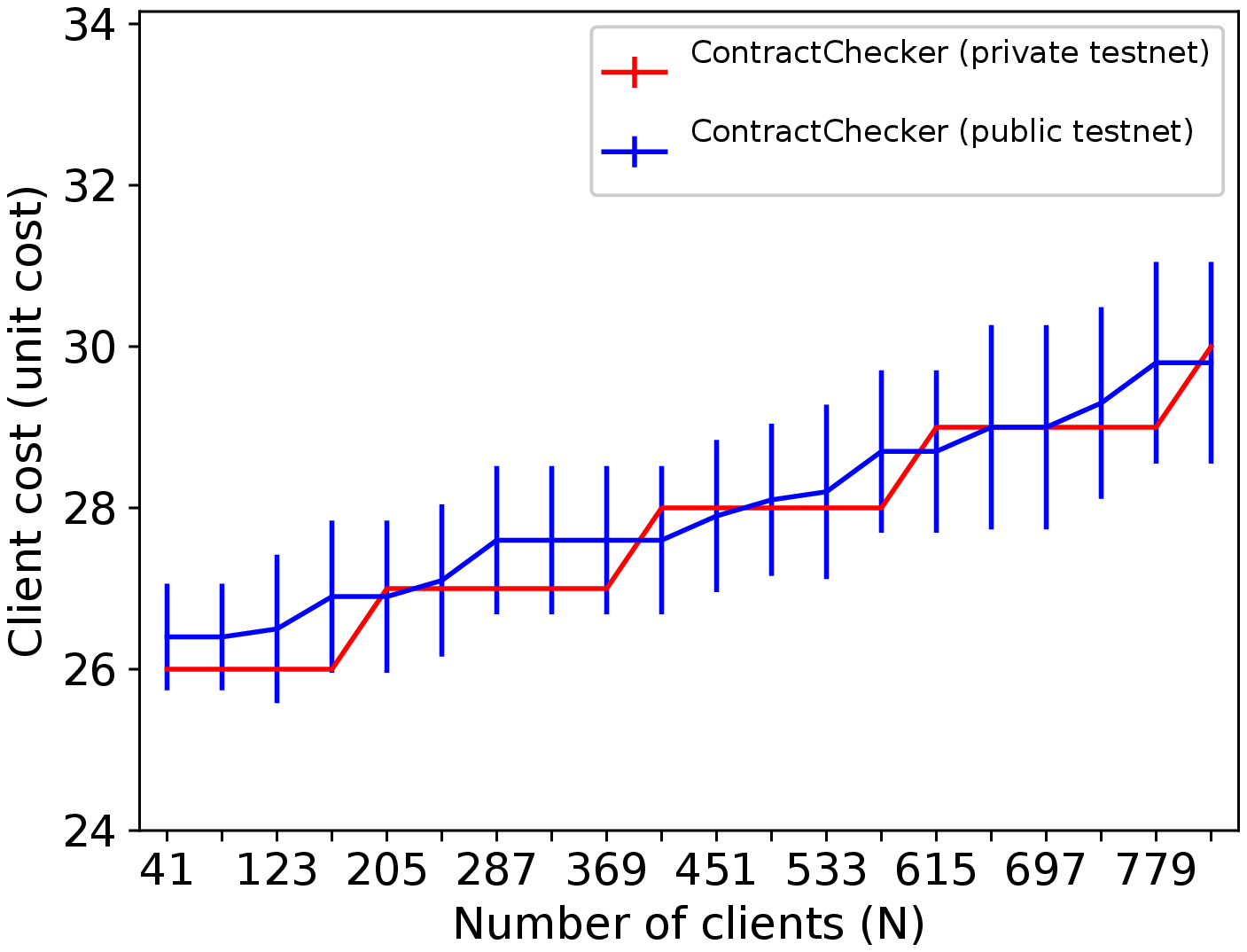}
    \label{exp:numclient:2}
}
\subfloat[With varying number of operations]{
   \includegraphics[width=0.225\textwidth]{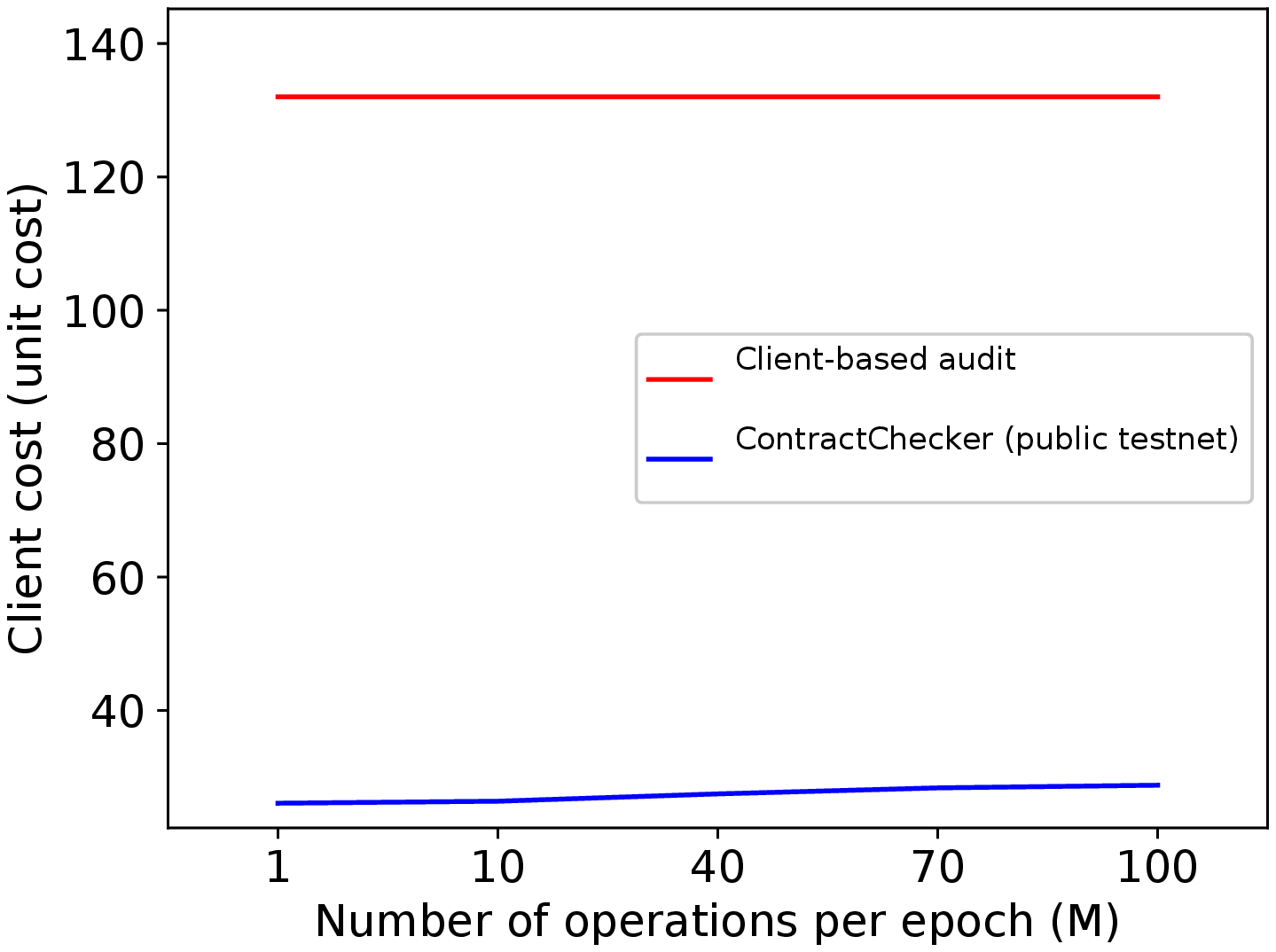}
    \label{exp:numops:1}
}
\subfloat[With varying number of operations]{
   \includegraphics[width=0.225\textwidth]{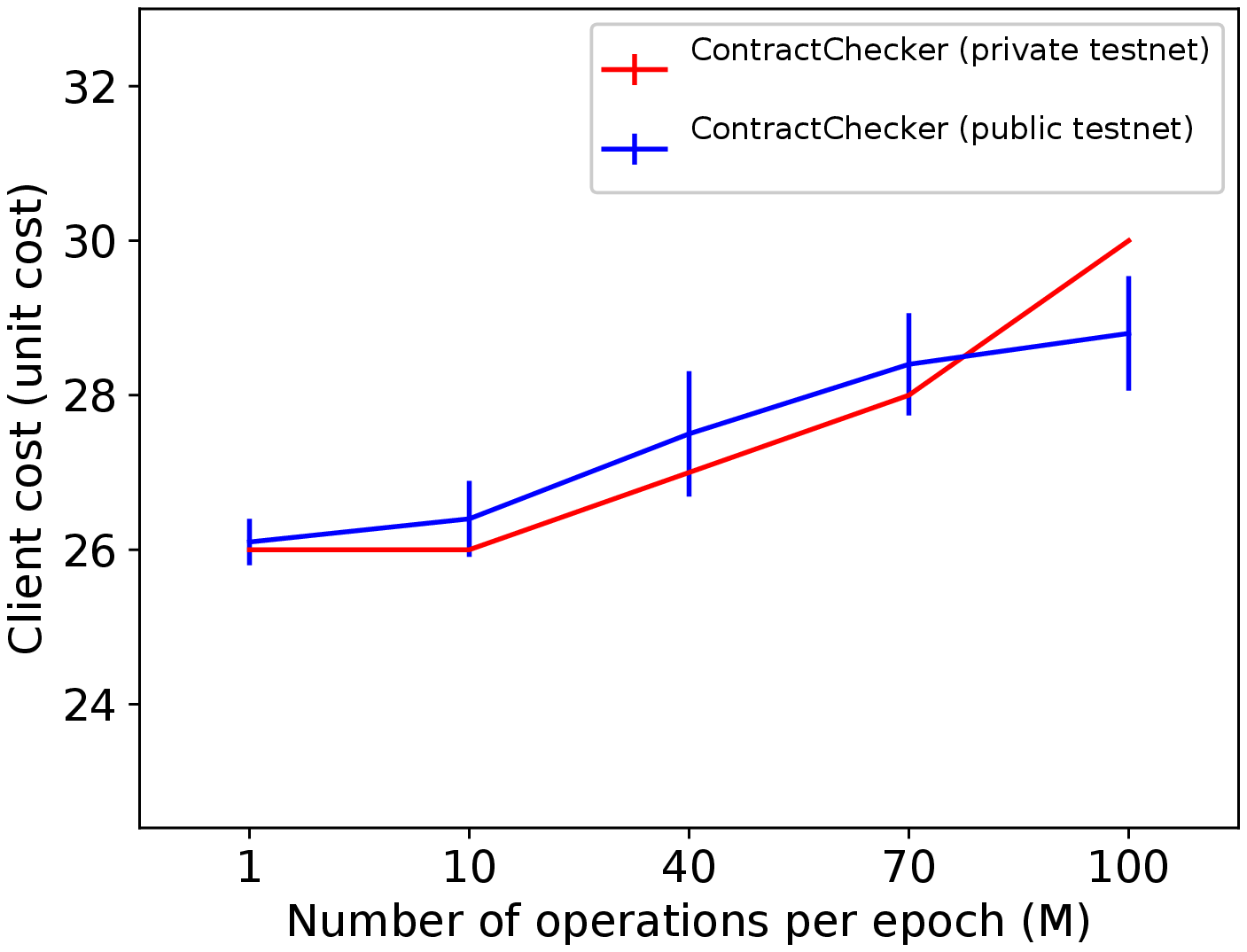}
    \label{exp:numops:2}
}
\end{center}
\caption{Client costs of ContractChecker}
\label{exp:capacityether}
\vspace{-0.10in}
\end{figure*} 

Based on the cost model above, we design experiments for client cost measurement. Our experiments focus on transaction-validation delay ($P$) and maximal number of operations ($N\cdot{}M$), w.r.t. the budgeted amount of Ethers. Based on the results, we compare ContractChecker's client cost with that of client-based auditing schemes.

We design experiments to evaluate the client cost of ContractChecker in comparison with client-based auditing schemes. From our cost analysis, the unit client cost of ContractChecker is $F_t/E$ (Equation~\ref{eqn:cc}) and that of client-based schemes is proportional to $N$ (Equation~\ref{eqn:clientaudit}). The source of extra cost is that ContractChecker needs to maintain a local operation for extra time and client-based schemes need to replicate a client's operation to all other clients. In this section, we aim at comparing the client costs of these two scheme paradigms using experiments.

Designing the experiments, we observe there are two factors affecting the cost, the number of clients (i.e., $N$) and the number of operations per epoch (i.e., $M$). Increasing $N$ and $M$, it will saturate the current block ingesting the operations in an epoch and delays the inclusion of all operations to next blocks. We thus design experiments to measure $F_t$ with varying $N$ and $M$. We consider the setting that $E$ is longer than $F_t$  such that the Blockchain will not permanently drop transactions. To measure $F_t$, we run $N$ clients to concurrently submit their attestation transactions to the Blockchain. The $N$ clients then proactively check the finality of the transactions. For simplicity, we use the maximal transaction fee to reduce the transaction pending time and exclude the case of dropped transactions. 
We record the time duration between the first transaction is submitted and the last transaction is finalized on Blockchain. 

{\bf Experiment setup}: We conduct experiments in a private Ethereum testnet and a public testnet. 
For the public testnet, we connect our ContractChecker protocol to the Ropsten testnet~\cite{me:ropsten}. 
For the private testnet, we run four Ethereum miners in an LAN network on campus. The machine specs is Intel 8-core i7-6820HK CPU of 2.70GHz, 32 GB RAM and 1 TB Disk. The default configuration for Ethereum (e.g., difficulty levels) is used. 
Note that our target metrics (e.g., transaction fee and Gas) are independent of network scale.
We deploy our smart-contract program written in Solidity to the Blockchain, by leveraging an online Python tool\footnote{\url{https://github.com/ConsenSys/ethjsonrpc}}.

We present the client costs with varying $N$ (number of clients) in Figure~\ref{exp:numclient:1} and Figure~\ref{exp:numclient:2}. In this experiment, we consider $M=1$ (each client submits one operation in one epoch) for the maximal scalability. From Figure~\ref{exp:numclient:1}, it can be seen that ContractChecker incurs much lower unit client cost than client-based schemes. With increasing number of clients, the client-based schemes incur a linearly increasing cost, while the ContractChecker's cost barely changes (it slightly increases as shown and analyzed in Figure~\ref{exp:numclient:2}). The reason behind is that with a new client, the client-based schemes require to replicate a given operation while ContractChecker does not replicate local operations to the new client. With an increasing $N$, the number of operations per epoch may increase, and that causes ContractChecker's client cost to increase --- with more operations in an epoch, it takes more blocks to include the operations, thus longer $F_t$. This can be seen in Figure~\ref{exp:numclient:2} more clearly. More specifically, with the public testnet, it takes one blocks to include all operations produced by up to $N=200$ clients, takes two blocks by up to $370$ clients, etc. The private Blockchain testnet incurs a slightly higher client cost than the public Blockchain\footnote{This result is a bit surprising to us, as the private Blockchain testnet is dedicated to ContractChecker while the public testnet is shared. We verify this is due to that in the private testnet a block can be utilized up to $80\%$ of specified gas limits, while the public testnet a block can be utilized $100\%$.}, and also has lower variance. Note that the increase rate of ContractChecker with $N$ is much slower than that of client-based schemes.

We present the client costs with varying $M$ (the average number of operations per epoch per client) in Figure~\ref{exp:numops:1}. In this experiment, we consider $N=132$ clients so that we can test up to $M=100$ operations. The result shows that ContractChecker's client cost is about one fifth of client-based auditing schemes' cost. With the increasing $M$, ContractChecker's cost slightly increases.

\subsubsection{Blockchain Availability and Transaction Fee}
\label{sec:availability}

\begin{figure}[!ht]
  \begin{center}
    \includegraphics[width=0.425\textwidth]{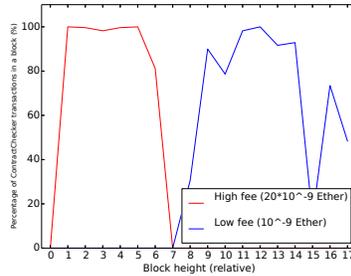}%
  \end{center}\vspace{-0.15in}
  \caption{Blockchain availability under varying transaction fee}\label{exp:txfee}
\end{figure}

In this experiment, our goal is to show the relationship between Blockchain write availability and transaction fee. We simultaneously run two Ethereum clients (on two separate machines), each of which drives 1500 transactions to the Ropsten testnet. The two clients respectively use high transaction fee ($20*10^{-9}$ Ether) and low fee ($1*10^{-9}$ Ether). All transactions are driven to the Blockchain at the time of one block (with the relative block height $0$) and we observe how and when transactions of different fees get included in the Blockchain. We repeatedly conduct 10 experiments and present the averaged result in Figure~\ref{exp:txfee}. In this figure, all 1500 transactions with high fee are included in the blocks before (or with smaller block height) the other 1500 transactions with low fee. More specifically, the 1500 transactions of high fee takes about 6 blocks to be included while the 1500 transactions of low fee takes up to 10 blocks to be included. This result shows that with higher transaction fee, transactions get included earlier and it is less likely to drop the transactions with lower fee.

\section{Conclusion}
This work presents ContractChecker, a lightweight consistency verification protocol based on public Blockchain. 
It presents a new approach by auditing the storage log on the Blockchain, and has advantages in saving the clients' cost and lower availability requirements.
ContractChecker is implemented in a middleware system federating cloud services, clients and the Blockchain.

\section*{Acknowledgement}
Yuzhe (Richard) Tang's work is supported by National Science Foundation under Grant CNS1815814 and a gift from Intel. Jianliang Xu's work is supported by Hong Kong RGC grants C6030-18GF, C1008-16G, and 12201018. The authors thank Yue Cheng for the discussion of this work in the early stage.


  \bibliographystyle{abbrv}
  \bibliography{ads,distrkvs,sc,bkc,sgx,cacheattacks,latex,txtbk,crypto,lsm,vc,diffpriv,odb,yuzhetang,yue}

\begin{thebibliography}{10}

\bibitem{me:s3}
Amazon s3: https://aws.amazon.com/s3.

\bibitem{me:ct}
Certificate transparency, https://tools.ietf.org/html/rfc6962.

\bibitem{me:dockerhub}
Docker hub: https://hub.docker.com/.

\bibitem{me:dropbox}
Dropbox: https://www.dropbox.com/.

\bibitem{me:eth}
Ethereum project: https://www.ethereum.org/.

\bibitem{me:contractattacks}
Ethereum smart contract best practices: Known attacks:
  https://consensys.github.io/smart-contract-best-practices/known\_attacks/.

\bibitem{me:github}
Github: https://github.com/.

\bibitem{me:icloudsync}
How to fix icloud sync in seconds,
  https://www.computerworld.com/article/2916476/apple-ios/how-to-fix-icloud-sync-in-seconds.html.

\bibitem{me:tr19cc}
Secure consistency verification for untrusted cloud storage by public
  blockchains.

\bibitem{me:ropsten}
Testnet ropsten (eth) blockchain explorer: https://ropsten.etherscan.io/.

\bibitem{docker_fast18}
A.~Anwar, M.~Mohamed, V.~Tarasov, M.~Littley, L.~Rupprecht, Y.~Cheng, N.~Zhao,
  D.~Skourtis, A.~S. Warke, H.~Ludwig, D.~Hildebrand, and A.~R. Butt.
\newblock Improving docker registry design based on production workload
  analysis.
\newblock In {\em 16th {USENIX} Conference on File and Storage Technologies
  ({FAST} 18)}, pages 265--278, Oakland, CA, 2018. {USENIX} Association.

\bibitem{DBLP:books/aw/BernsteinHG87}
P.~A. Bernstein, V.~Hadzilacos, and N.~Goodman.
\newblock {\em Concurrency Control and Recovery in Database Systems}.
\newblock Addison-Wesley, 1987.

\bibitem{DBLP:conf/cloud/CooperSTRS10}
B.~F. Cooper, A.~Silberstein, E.~Tam, R.~Ramakrishnan, and R.~Sears.
\newblock Benchmarking cloud serving systems with ycsb.
\newblock In {\em SoCC}, pages 143--154, 2010.

\bibitem{DBLP:journals/tocs/CorbettDEFFFGGHHHKKLLMMNQRRSSTWW13}
J.~C. Corbett, J.~Dean, M.~Epstein, A.~Fikes, C.~Frost, J.~J. Furman,
  S.~Ghemawat, A.~Gubarev, C.~Heiser, P.~Hochschild, W.~C. Hsieh, S.~Kanthak,
  E.~Kogan, H.~Li, A.~Lloyd, S.~Melnik, D.~Mwaura, D.~Nagle, S.~Quinlan,
  R.~Rao, L.~Rolig, Y.~Saito, M.~Szymaniak, C.~Taylor, R.~Wang, and
  D.~Woodford.
\newblock Spanner: Google's globally distributed database.
\newblock {\em {ACM} Trans. Comput. Syst.}, 31(3):8, 2013.

\bibitem{DBLP:journals/toplas/HerlihyW90}
M.~Herlihy and J.~M. Wing.
\newblock Linearizability: {A} correctness condition for concurrent objects.
\newblock {\em {ACM} Trans. Program. Lang. Syst.}, 12(3):463--492, 1990.

\bibitem{DBLP:conf/sp/KimL15}
B.~H. Kim and D.~Lie.
\newblock Caelus: Verifying the consistency of cloud services with
  battery-powered devices.
\newblock In {\em 2015 {IEEE} Symposium on Security and Privacy, {SP} 2015, San
  Jose, CA, USA, May 17-21, 2015}, pages 880--896. {IEEE} Computer Society,
  2015.

\bibitem{DBLP:conf/osdi/LiKMS04}
J.~Li, M.~N. Krohn, D.~Mazi{\`e}res, and D.~Shasha.
\newblock Secure untrusted data repository (sundr).
\newblock In {\em OSDI}, pages 121--136, 2004.

\bibitem{DBLP:conf/podc/MazieresS02}
D.~Mazi{\`{e}}res and D.~Shasha.
\newblock Building secure file systems out of byantine storage.
\newblock In {\em Proceedings of the Twenty-First Annual {ACM} Symposium on
  Principles of Distributed Computing, {PODC} 2002, Monterey, California, USA,
  July 21-24, 2002}, pages 108--117, 2002.

\bibitem{DBLP:conf/uss/MelaraBBFF15}
M.~S. Melara, A.~Blankstein, J.~Bonneau, E.~W. Felten, and M.~J. Freedman.
\newblock {CONIKS:} bringing key transparency to end users.
\newblock In J.~Jung and T.~Holz, editors, {\em 24th {USENIX} Security
  Symposium, {USENIX} Security 15, Washington, D.C., USA, August 12-14, 2015.},
  pages 383--398. {USENIX} Association, 2015.

\bibitem{DBLP:conf/usenix/PopaLMWZ11}
R.~A. Popa, J.~R. Lorch, D.~Molnar, H.~J. Wang, and L.~Zhuang.
\newblock Enabling security in cloud storage slas with cloudproof.
\newblock In J.~Nieh and C.~A. Waldspurger, editors, {\em 2011 {USENIX} Annual
  Technical Conference, Portland, OR, USA, June 15-17, 2011}. {USENIX}
  Association, 2011.

\bibitem{DBLP:journals/cacm/Terry13}
D.~Terry.
\newblock Replicated data consistency explained through baseball.
\newblock {\em Commun. {ACM}}, 56(12):82--89, 2013.

\bibitem{DBLP:conf/sosp/TerryPKBAA13}
D.~B. Terry, V.~Prabhakaran, R.~Kotla, M.~Balakrishnan, M.~K. Aguilera, and
  H.~Abu{-}Libdeh.
\newblock Consistency-based service level agreements for cloud storage.
\newblock In M.~Kaminsky and M.~Dahlin, editors, {\em {ACM} {SIGOPS} 24th
  Symposium on Operating Systems Principles, {SOSP} '13, Farmington, PA, USA,
  November 3-6, 2013}, pages 309--324. {ACM}, 2013.

\bibitem{DBLP:conf/sp/TomescuD17}
A.~Tomescu and S.~Devadas.
\newblock Catena: Efficient non-equivocation via bitcoin.
\newblock In {\em 2017 {IEEE} Symposium on Security and Privacy, {SP} 2017, San
  Jose, CA, USA, May 22-26, 2017}, pages 393--409. {IEEE} Computer Society,
  2017.

\bibitem{DBLP:conf/uss/Torres-AriasACC16}
S.~Torres{-}Arias, A.~K. Ammula, R.~Curtmola, and J.~Cappos.
\newblock On omitting commits and committing omissions: Preventing git metadata
  tampering that (re)introduces software vulnerabilities.
\newblock In T.~Holz and S.~Savage, editors, {\em 25th {USENIX} Security
  Symposium, {USENIX} Security 16, Austin, TX, USA, August 10-12, 2016.}, pages
  379--395. {USENIX} Association, 2016.

\bibitem{DBLP:conf/sp/ZyskindNP15}
G.~Zyskind, O.~Nathan, and A.~Pentland.
\newblock Decentralizing privacy: Using blockchain to protect personal data.
\newblock In {\em 2015 {IEEE} Symposium on Security and Privacy Workshops,
  {SPW} 2015, San Jose, CA, USA, May 21-22, 2015}, pages 180--184. {IEEE}
  Computer Society, 2015.

\end{thebibliography}
 \appendix

\end{document}